\documentstyle[12pt,epsf]{article}

\def\doublespace{ \renewcommand{\baselinestretch}{1.7} \large\normalsize }
\def\singlespace{ \renewcommand{\baselinestretch}{1} \large\normalsize }

\begin{document}

\date{February 12, 1996}

\def\spose#1{\hbox to 0pt{#1\hss}}
\def\ltapprox{\mathrel{\spose{\lower 3pt\hbox{$\mathchar"218$}}
 \raise 2.0pt\hbox{$\mathchar"13C$}}}
\def\gtapprox{\mathrel{\spose{\lower 3pt\hbox{$\mathchar"218$}}
 \raise 2.0pt\hbox{$\mathchar"13E$}}}
\def\inapprox{\mathrel{\spose{\lower 3pt\hbox{$\mathchar"218$}}
 \raise 2.0pt\hbox{$\mathchar"232$}}}

\font\sevenrm  = cmr7 
\def\fancyplus{\hbox{+\kern-6.65pt\lower3.2pt\hbox{\sevenrm --}%
\kern-4pt\raise4.6pt\hbox{\sevenrm --}%
\kern-9pt\raise0.65pt\hbox{$\tiny\vdash$}%
\kern-2pt\raise0.65pt\hbox{$\tiny\dashv$}}}
\def\fancycross{\hbox{$\times$\kern-9.8pt\raise3.6pt\hbox{$\tiny \times$}%
\kern-1.2pt\raise3.6pt\hbox{$\tiny \times$}%
\kern-10.5pt\lower1.0pt\hbox{$\tiny \times$}%
\kern-1.2pt\lower1.0pt\hbox{$\tiny \times$}}}
\def\fancysquare{%
\hbox{$\small\Box$\kern-9.6pt\raise7.1pt\hbox{$\vpt\backslash$}%
\kern+3.9pt\raise7.1pt\hbox{$\vpt /$}%
\kern-11.5pt\lower3.0pt\hbox{$\vpt /$}%
\kern+3.9pt\lower3.0pt\hbox{$\vpt\backslash$}}}
\def\fancydiamond{\hbox{$\diamond$\kern-4.25pt\lower3.8pt\hbox{$\vpt\vert$}%
\kern-2.25pt\raise7pt\hbox{$\vpt\vert$}%
\kern-7.05pt\raise1.57pt\hbox{\vpt --}%
\kern+5.25pt\raise1.57pt\hbox{\vpt --}}}

\title{\vspace*{-2cm} Asymptotic Scaling in the\break
           Two-Dimensional $SU(3)$ $\sigma$-Model\break
           at Correlation Length $4 \times 10^5$}

\author{
  {\small Gustavo Mana}                   \\[-0.2cm]
  {\small\it Department of Physics}       \\[-0.2cm]
  {\small\it New York University}         \\[-0.2cm]
  {\small\it 4 Washington Place}          \\[-0.2cm]
  {\small\it New York, NY 10003 USA}      \\[-0.2cm]
  {\small Internet: {\tt MANA@MAFALDA.PHYSICS.NYU.EDU}}        \\[-0.2cm]
  \\[-4mm]  \and
  {\small Andrea Pelissetto}              \\[-0.2cm]
  {\small\it Dipartimento di Fisica and INFN -- Sezione di Pisa}  \\[-0.2cm]
  {\small\it Universit\`a degli Studi di Pisa}  \\[-0.2cm]
  {\small\it I-56100 Pisa, ITALIA}              \\[-0.2cm]
  {\small Internet: {\tt PELISSET@SUNTHPI1.DIFI.UNIPI.IT}}  \\[-0.2cm]
  \\[-4mm]  \and
  {\small Alan D. Sokal}                  \\[-0.2cm]
  {\small\it Department of Physics}       \\[-0.2cm]
  {\small\it New York University}         \\[-0.2cm]
  {\small\it 4 Washington Place}          \\[-0.2cm]
  {\small\it New York, NY 10003 USA}      \\[-0.2cm]
  {\small Internet: {\tt SOKAL@NYU.EDU}}        \\[-0.2cm]
  {\protect\makebox[5in]{\quad}}  
  \\
}

\maketitle
\thispagestyle{empty}   

\doublespace

\begin{abstract}
We carry out a high-precision simulation of the
two-dimensional $SU(3)$ principal chiral model
at correlation lengths $\xi$ up to $\approx\! 4 \times 10^5$,
using a multi-grid Monte Carlo (MGMC) algorithm.
We extrapolate
the finite-volume Monte Carlo data to infinite volume
using finite-size-scaling theory,
and we discuss carefully the systematic and statistical errors
in this extrapolation.
We then compare the extrapolated data to the renormalization-group predictions.
For $\xi \gtapprox 10^3$ we observe good asymptotic scaling in the 
bare coupling; at $\xi \approx 4 \times 10^5$ the nonperturbative 
constant is within 2--3\% of its predicted limiting value.
\end{abstract}

\noindent
{\bf PACS number(s):}  11.10.Gh, 11.15.Ha, 12.38.Gc, 05.70.Jk

\clearpage

\newcommand{\be}{\begin{equation}}
\newcommand{\ee}{\end{equation}}
\newcommand{\<}{\langle}
\renewcommand{\>}{\rangle}
\newcommand{\para}{\|}
\renewcommand{\perp}{\bot}

\def\half{ {{1 \over 2 }}}
\def\smfrac#1#2{{\textstyle\frac{#1}{#2}}}
\def\smhalf{ {\smfrac{1}{2}} }
\def\scra{{\cal A}}
\def\scrc{{\cal C}}
\def\scre{{\cal E}}
\def\scrf{{\cal F}}
\def\scrh{{\cal H}}
\def\scrm{{\cal M}}
\newcommand{\scrmvec}{\vec{\cal M}}
\def\scro{{\cal O}}
\def\scrp{{\cal P}}
\def\scrr{{\cal R}}
\def\scrs{{\cal S}}
\def\scrt{{\cal T}}
\def\ttens{{\stackrel{\leftrightarrow}{T}}}
\def\scrttens{{\stackrel{\leftrightarrow}{\cal T}}}
\def\scrv{{\cal V}}
\def\scrw{{\cal W}}
\def\scry{{\cal Y}}
\def\tauss{\tau_{int,\,\scrm^2}}
\def\taux{\tau_{int,\,{\cal M}^2}}
\newcommand{\taum}{\tau_{int,\,\vec{\cal M}}}
\def\taue{\tau_{int,\,{\cal E}}}
\newcommand{\imag}{\mathop{\rm Im}\nolimits}
\newcommand{\real}{\mathop{\rm Re}\nolimits}
\newcommand{\tr}{\mathop{\rm tr}\nolimits}
\newcommand{\sgn}{\mathop{\rm sgn}\nolimits}
\newcommand{\codim}{\mathop{\rm codim}\nolimits}
\def\textprime{{${}^\prime$}}
\newcommand{\longto}{\longrightarrow}
\def\var{ \hbox{var} }
\newcommand{\gtilde}{ {\widetilde{G}} }
\newcommand{\USp}{ \hbox{\it USp} }
\newcommand{\CP}{ \hbox{\it CP\/} }
\newcommand{\QP}{ \hbox{\it QP\/} }
\def\hboxscript#1{ {\hbox{\scriptsize\em #1}} }

\newcommand{\plotdot}{\makebox(0,0){$\bullet$}}
\newcommand{\plotsmalldot}{\makebox(0,0){{\footnotesize $\bullet$}}}

\def\bsigma{\mbox{\protect\boldmath $\sigma$}}
\def\btau{\mbox{\protect\boldmath $\tau$}}
\def\br{{\bf r}}

\newcommand{\reff}[1]{(\ref{#1})}

\font\specialroman=msym10 scaled\magstep1  
\font\sevenspecialroman=msym7              
\def\zed{\hbox{\specialroman Z}}
\def\szed{\hbox{\sevenspecialroman Z}}
\def\R{\hbox{\specialroman R}}
\def\sR{\hbox{\sevenspecialroman R}}
\def\N{\hbox{\specialroman N}}
\def\C{\hbox{\specialroman C}}
\def\Q{\hbox{\specialroman Q}}
\renewcommand{\emptyset}{\hbox{\specialroman ?}}

\newtheorem{theorem}{Theorem}[section]
\newtheorem{corollary}[theorem]{Corollary}
\newtheorem{lemma}[theorem]{Lemma}
\def\proof{\bigskip\par\noindent{\sc Proof.\ }}
\def\qed{\hbox{\hskip 6pt\vrule width6pt height7pt depth1pt \hskip1pt}\bigskip}

\newenvironment{sarray}{
          \textfont0=\scriptfont0
          \scriptfont0=\scriptscriptfont0
          \textfont1=\scriptfont1
          \scriptfont1=\scriptscriptfont1
          \textfont2=\scriptfont2
          \scriptfont2=\scriptscriptfont2
          \textfont3=\scriptfont3
          \scriptfont3=\scriptscriptfont3
        \renewcommand{\arraystretch}{0.7}
        \begin{array}{l}}{\end{array}}

\newenvironment{scarray}{
          \textfont0=\scriptfont0
          \scriptfont0=\scriptscriptfont0
          \textfont1=\scriptfont1
          \scriptfont1=\scriptscriptfont1
          \textfont2=\scriptfont2
          \scriptfont2=\scriptscriptfont2
          \textfont3=\scriptfont3
          \scriptfont3=\scriptscriptfont3
        \renewcommand{\arraystretch}{0.7}
        \begin{array}{c}}{\end{array}}

A key tenet of modern elementary-particle physics is the asymptotic freedom
of four-dimensional nonabelian gauge theories \cite{AF_4d_gauge}.
However, the nonperturbative validity of asymptotic freedom has been
questioned \cite{Pat-Seil};
and numerical studies of lattice gauge theory
have thus far failed to detect asymptotic scaling in the bare coupling
\cite{4d_gauge_asymptotic_scaling}.
Even in the simpler case of two-dimensional nonlinear $\sigma$-models
\cite{AF_conventional},
numerical simulations at correlation lengths $\xi \sim 10$--100
have often shown discrepancies of order 10--50\% 
from asymptotic scaling.
In a recent paper \cite{o3_scaling_prl}
we employed a finite-size-scaling extrapolation method
\cite{Luscher_91,Kim_93,fss_greedy,related_work_explanation}
to carry simulations in the $O(3)$ $\sigma$-model to
correlation lengths $\xi \approx  10^5$;
the discrepancy from asymptotic scaling decreased from $\approx\! 25\%$
to $\approx\! 4\%$.
In the present Letter we apply a similar technique to the
$SU(3)$ principal chiral model,
reaching correlation lengths $\xi \approx 4 \times 10^5$
with errors $\ltapprox 2\%$.
For $\xi \gtapprox 10^3$ we observe good asymptotic scaling in the
bare parameter $\beta$; moreover, at $\xi \approx 4 \times 10^5$ the 
nonperturbative ratio $\xi_{observed}/\xi_{theor,3-loop}$
is within 2--3\% of the predicted limiting value.

We study the lattice $\sigma$-model taking values in the group $SU(N)$, 
with nearest-neighbor action
$\scrh (U) = - \beta \sum \real\tr(U_{x}^\dagger \, U_{y})$.
Perturbative renormalization-group computations predict
that the infinite-volume correlation lengths $\xi^{(exp)}$ and $\xi^{(2)}$
\cite{correlation_lengths}
behave as
\be
\xi^\#(\beta)    \;=\;   C_{\xi^\#} \, e^{4\pi\beta/N} \,
       \left( {4\pi\beta \over N} \right) ^{\! -1/2}   \,
       \left[ 1 + {a_1 \over \beta} + {a_2 \over \beta^2} + \cdots \right]
 \label{xi_predicted}
\ee
as $\beta \to \infty$.
Three-loop perturbation theory yields \cite{Rossi_94}
\begin{eqnarray}
   a_1   & = &   - 0.121019 N +
                   0.725848 N^{-1} - 1.178097 N^{-3}  \;.
\end{eqnarray}
The nonperturbative constant $C_{\xi^{(exp)}}$ has been computed
using the thermodynamic Bethe Ansatz \cite{Balog_92}:
\be
   C_{\xi^{(exp)}}
   \;=\;
   {\sqrt{e} \over  16 \sqrt{\pi}} \,
   {\pi/N    \over  \sin(\pi/N)}   \,
   \exp\!\left( -\pi {N^2 - 2  \over  2N^2} \right)
   \;.
 \label{exact_Cxi}
\ee
The nonperturbative constant $C_{\xi^{(2)}}$ is unknown,
but Monte Carlo studies indicate that $C_{\xi^{(2)}}/C_{\xi^{(exp)}}$
lies between $\approx\! 0.985$ and 1 for all $N \ge 2$
\cite{constant_SUN}; for $N=3$ it is $0.987 \pm 0.002$ \cite{Rossi_94}.
Monte Carlo studies
\cite{Dagotto_87:SU(3),Hasenbusch-Meyer_PRL,Horgan_LAT93,Rossi_94}
of the $SU(3)$ model up to $\xi \approx 35$ have failed to observe
asymptotic scaling \reff{xi_predicted}; the discrepancy from
\reff{xi_predicted}--\reff{exact_Cxi} is of order 10--20\%.

Our extrapolation method \cite{fss_greedy}
is based on the finite-size-scaling (FSS) Ansatz
\be 
   {\scro(\beta,sL) \over \scro(\beta,L)}   \;=\;
   F_{\scro} \Bigl( \xi(\beta,L)/L \,;\, s \Bigr)
   \,+\,  O \Bigl( \xi^{-\omega}, L^{-\omega} \Bigr)
   \;,
 \label{eq2}
\ee
where $\scro$ is any long-distance observable,
$s$ is a fixed scale factor (here $s=2$),
$L$ is the linear lattice size,
$F_{\scro}$ is a universal function,
and $\omega$ is a correction-to-scaling exponent.
We make Monte Carlo runs at numerous pairs $(\beta,L)$ and $(\beta,sL)$;
we then plot $\scro(\beta,sL) / \scro(\beta,L)$ versus $\xi(\beta,L)/L$,
using those points satisfying both $\xi(\beta,L) \ge$ some value $\xi_{min}$
and $L \ge$ some value $L_{min}$.
If all these points fall with good accuracy on a single curve,
we choose a smooth fitting function $F_{\scro}$.
Then, using the functions $F_\xi$ and $F_\scro$,
we extrapolate the pair $(\xi,\scro)$ successively from
$L \to sL \to s^2 L \to \ldots \to \infty$.
See \cite{fss_greedy} for how to calculate statistical error bars
on the extrapolated values.

We have chosen to use functions $F_{\scro}$ of the form
\be
   F_\scro(x)   \;=\;
   1 + a_1 e^{-1/x} + a_2 e^{-2/x} + \ldots + a_n e^{-n/x}   \;.
 \label{eq3}
\ee
We increase $n$ until the $\chi^2$ of the fit becomes essentially constant;
the resulting $\chi^2$ value provides a check on the
systematic errors arising from corrections to scaling and/or
from inadequacies of the form \reff{eq3}.
The discrepancies between the extrapolated values from different
$L$ at the same $\beta$ can also be subjected to a $\chi^2$ test.
Further details on the method can be found in \cite{fss_greedy,o3_scaling_prl}.

We simulated the two-dimensional $SU(3)$ $\sigma$-model
using an $XY$-embedding multi-grid Monte Carlo (MGMC) algorithm
\cite{Mendes_LAT95}.
We ran on lattices $L=8,16,32,64,$ 
$128,256$ at 184 different pairs $(\beta,L)$
in the range $1.65 \le \beta \le 4.35$
(corresponding to $5 \ltapprox \xi_\infty \ltapprox 4 \times 10^5$).
Each run was between $4 \times 10^5$ and $5 \times 10^6$ iterations,
and the total CPU time was one year on a Cray C-90 \cite{CPU_time}.
The raw data will appear in \cite{MGMC_SU3}.

\begin{figure}
\vspace*{0cm} \hspace*{-0cm}
\begin{center}
\epsfxsize = 0.9\textwidth
\leavevmode\epsffile{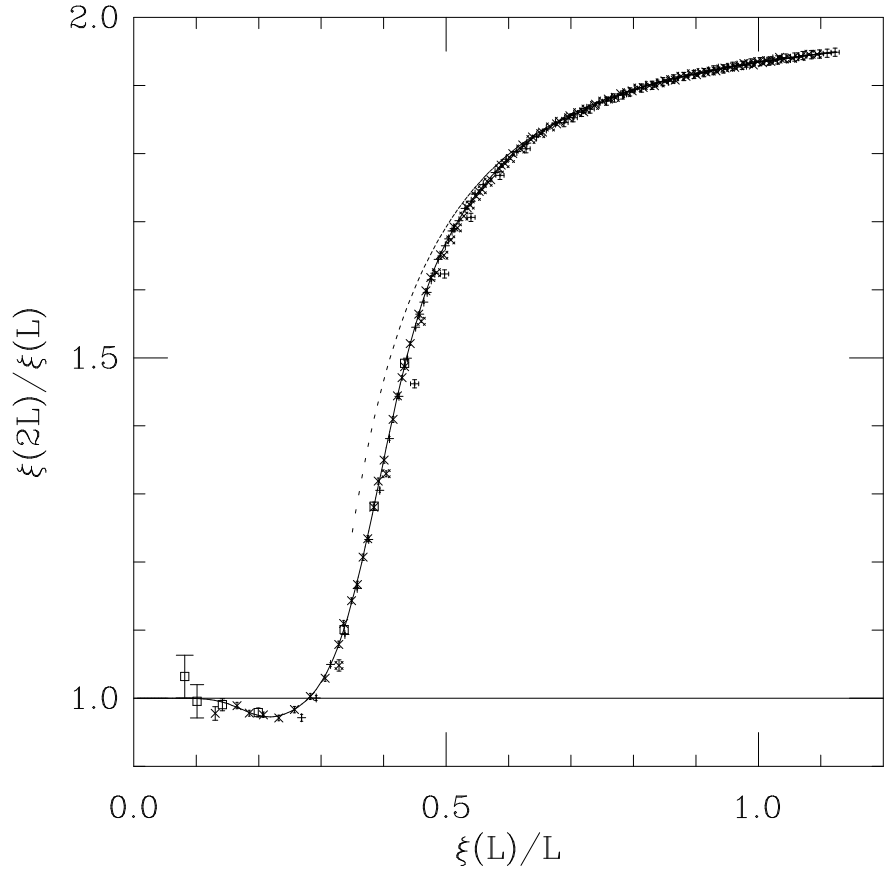}
\end{center}
\vspace*{-2cm}
\caption{
   $\xi(\beta,2L)/\xi(\beta,L)$ versus $\xi(\beta,L)/L$.
   Symbols indicate $L=8$ ($\protect\fancyplus$), 16 ($\protect\fancycross$),
   32 ($+$), 64 ($\times$), 128 ($\Box$).
   Error bars are one standard deviation.
   Solid curve is a thirteenth-order fit in \protect\reff{eq3},
   with $x_{min} = (\infty,0.90,0.65,0.14,0)$ for
   $L = (8,16,32,64,128)$.
   Dashed curve is the perturbative prediction (\protect\ref{PT_Fxi}).
}
\label{fig_su3_xiplot}
\end{figure}

Our FSS data 
cover the range $0.08 \ltapprox x \equiv \xi(L)/L \ltapprox 1.12$,
and we found tentatively that for $\scro = \xi$
a thirteenth-order fit \reff{eq3} is indicated:
see Table~\ref{chisquared_xi_versus_n}.
There are significant corrections to scaling 
in the regions $x \ltapprox 0.84$ (resp.\ 0.64, 0.52, 0.14)
when $L=8$ (resp.\ 16, 32, 64):
see the deviations plotted in Figure~\ref{deviazioni_curva}.
We therefore investigated systematically the $\chi^2$ of the fits,
allowing different cuts in $x$ for different values of $L$:
see again Table~\ref{chisquared_xi_versus_n}.
A reasonable $\chi^2$ is obtained when $n \ge 13$ and
$x_{min} \ge (0.80,0.70,0.60,0.14,0)$ for $L=(8,16,32,64,128)$.
Our preferred fit is $n=13$ and $x_{min} = (\infty,0.90,0.65,0.14,0)$:
see Figure \ref{fig_su3_xiplot}, where we compare also with the
perturbative prediction
\be
   F_\xi(x;s)   \;=\;
   s \left[ 1 \,-\,  {a w_0 \log s \over 2} x^{-2}
       \,-\,  a^2 \Biggl( {w_1 \log s \over 2} + {w_0^2 \log^2 s \over 8}
                  \Biggr) x^{-4}
       \,+\,  O(x^{-6}) \right]
  \label{PT_Fxi}
\ee
valid for $x \gg 1$,
where $a = 2N/(N^2 - 1)$, 
$w_0 = N/(8 \pi)$ and $w_1 = N^2/(128 \pi^2)$.

The extrapolated values $\xi_\infty^{(2)}$ from different lattice sizes
at the same $\beta$ are consistent within statistical errors:
only one of the 58 $\beta$ values has a $\chi^2$
too large at the 5\% level;
and summing all $\beta$ values we have $\chi^2 = 64.28$ 
(103 DF, level = 99.9\%).

In Table~\ref{table_su3_xi} we show the extrapolated values $\xi_\infty^{(2)}$
from our preferred fit and some alternative fits.
The deviations between the different fits
(if larger than the statistical errors)
can serve as a rough estimate of the remaining systematic errors due
to corrections to scaling.
The statistical errors in our preferred fit are of order
0.5\% (resp.\ 0.9\%, 1.1\%, 1.3\%, 1.5\%)
at $\xi_\infty \approx 10^2$ (resp.\ $10^3$, $10^4$, $10^5$, $4 \times 10^5$),
and the systematic errors are of the same order or smaller.
The statistical errors at different $\beta$ are strongly positively correlated.

\begin{figure}
\vspace*{0cm} \hspace*{-0cm}
\begin{center}
\epsfxsize = 0.9\textwidth
\leavevmode\epsffile{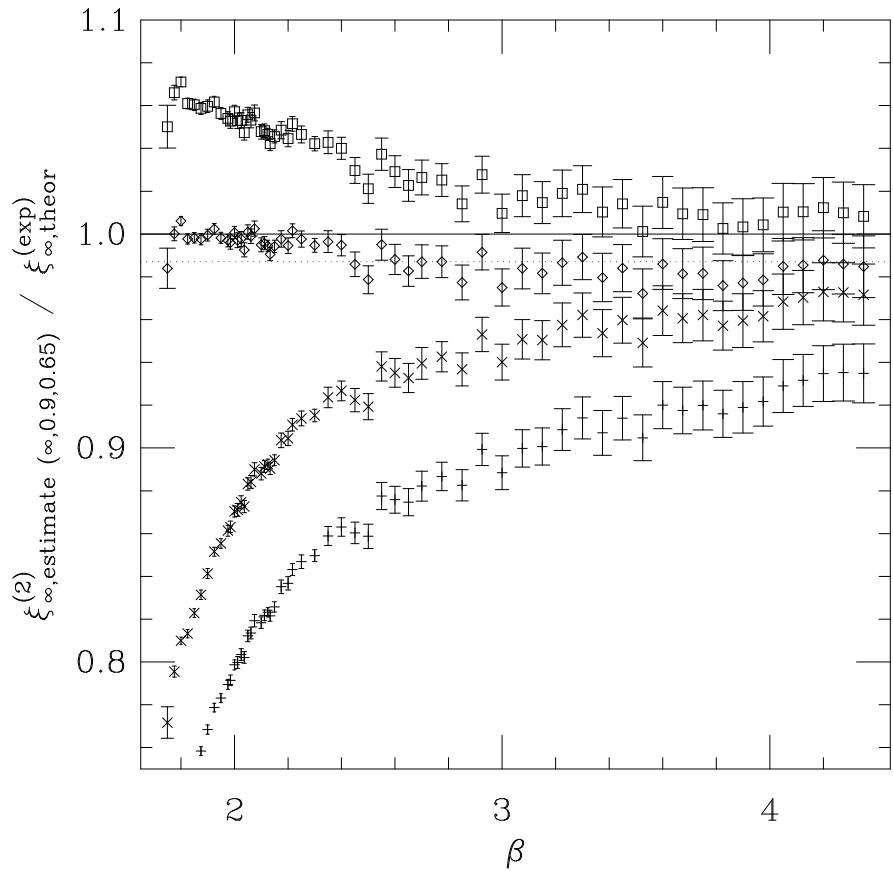}
\end{center}
\vspace*{-2cm}
\caption{
   $\xi^{(2)}_{\infty,estimate\,(\infty,0.90,0.65)}/\xi^{(exp)}_{\infty,theor}$
   versus $\beta$.
   Error bars are one standard deviation (statistical error only).
   There are four versions of $\xi^{(exp)}_{\infty,theor}$:
   standard perturbation theory in $1/\beta$ gives points
   $+$ (2-loop) and $\times$ (3-loop);
   ``improved'' perturbation theory in $1-E$ gives points
   $\Box$ (2-loop) and $\Diamond$ (3-loop). Dotted line is 
   the Monte Carlo prediction 
   $C_{\xi^{(2)}}/C_{\xi^{(exp)}} = 0.987 \pm 0.002$ \protect\cite{Rossi_94}.
}
\label{fig_su3_3loop}
\end{figure}

In Figure~\ref{fig_su3_3loop} (points $+$ and $\times$)
we plot $\xi^{(2)}_{\infty,estimate\,(\infty,0.90,0.65)}$
divided by the two-loop and three-loop predictions
\reff{xi_predicted}--\reff{exact_Cxi} for $\xi^{(exp)}$.
The discrepancy from three-loop asymptotic scaling,
which is $\approx 13\%$ at $\beta=2.0$ ($\xi_{\infty} \approx 25$),
decreases to 2--3\% at $\beta=4.35$ 
($\xi_{\infty} \approx 3.7 \times 10^5$).
For $\beta \gtapprox 2.2$ ($\xi_{\infty} \gtapprox 60$)
our data are consistent with convergence to 
a limiting value $C_{\xi^{(2)}}/C_{\xi^{(exp)}} \approx 0.99$--1
with the expected $1/\beta^2$ corrections.

We can also try an ``improved expansion parameter''
\cite{improved_exp,Rossi_94}
based on the energy $E = N^{-1} \< \real\tr(U_0^\dagger \, U_1) \>$.
First we invert the perturbative expansion \cite{Rossi_94}
\be
   E(\beta) \;=\;  1 \,-\, {N^{2}-1 \over 4N\beta}  \, \left[ 1
                   \,+\, {N^2-2 \over 16N\beta}
          \,+\, {0.0756 - 0.0634\,{N^2} + 0.01743\,{N^4}
                   \over  {N^2}\beta^2}
                   \,+\, O(1/\beta^3) \right]
\ee
and substitute into \reff{xi_predicted};
this gives a prediction for $\xi$ as a function of $1-E$.
For $E$ we use the value measured on the largest lattice (which is usually 
$L = 128$);
the statistical errors and finite-size corrections on $E$
are less than $5\times 10^{-4}$,
 and they induce an error less than $0.85\%$
on the predicted $\xi_\infty$ (less than 0.55\% for $\beta \geq 2.2$).
The corresponding observed/predicted ratios are also shown in
Figure~\ref{fig_su3_3loop} (points $\Box$ and $\Diamond$).
The ``improved'' 3-loop prediction is extremely flat, and again
indicates a limiting value $\approx 0.99$.

Further discussion of the conceptual basis of our analysis
can be found in \cite{o3_scaling_prl}.
Details of this work, including an analysis of the susceptibility $\chi$,
will appear elsewhere \cite{MGMC_SU3}.

\bigskip

A.P. would like to thank NYU for hospitality while some of this
work was being carried out.
These computations were performed on the
Cray C-90 at the Pittsburgh Supercomputing Center
and on the IBM SP2 cluster at the Cornell Theory Center.
The authors' research was supported by
NSF grants DMS-9200719 and PHY-9520978.

\clearpage

\begin{figure}
\vspace*{0cm} \hspace*{-0cm}
\begin{center}
\epsfxsize = 0.9\textwidth
\leavevmode\epsffile{su3_fig1.ps}
\end{center}
\vspace*{-2cm}
\caption{
   Deviation of points from fit to $F_\xi$ with $s=2$, $n=13$,
   $x_{min} = (\infty,\infty,\infty,0.14,0)$.
   Symbols indicate $L=8$ ($\protect\fancyplus$), 16 ($\protect\fancycross$),
   32 ($+$).
   Error bars are one standard deviation.
   Curves near zero indicate statistical error bars
   ($\pm$ one standard deviation) on the function $F_\xi(x)$.
}
\label{deviazioni_curva}
\end{figure}

\clearpage

\clearpage

\clearpage

\singlespace
\protect\small
\begin{table}
\begin{center}
\begin{tabular}{|c||c|c|c|c|c|}
\hline
$x_{min}$  &  $n=11$  & $n=12$  & $n=13$  & $n=14$ & $n=15$  \\ \hline \hline
(0.50,0.40,0)                & \rm 180  718.80  & \rm 179  626.60  & \rm 178  560.20  & \rm 177  558.60  & \rm 176  558.30  \\
                             & \rm 3.99   0.0\% & \rm 3.50   0.0\% & \rm 3.15   0.0\% & \rm 3.16   0.0\% & \rm 3.17   0.0\% \\ \hline
($\infty$,0.40,0)            & \rm 154  673.80  & \rm 153  566.30  & \rm 152  533.00  & \rm 151  532.10  & \rm 150  531.80  \\
                             & \rm 4.38   0.0\% & \rm 3.70   0.0\% & \rm 3.51   0.0\% & \rm 3.52   0.0\% & \rm 3.55   0.0\% \\ \hline
($\infty$,$\infty$,0)        & \rm 108  236.00  & \rm 107  172.40  & \rm 106  154.80  & \rm 105  154.70  & \rm 104  153.40  \\
                             & \rm 2.19   0.0\% & \rm 1.61   0.0\% & \rm 1.46   0.1\% & \rm 1.47   0.1\% & \rm 1.48   0.1\% \\ \hline
(0.70,0.55,0.45)             & \rm 162  288.30  & \rm 161  219.20  & \rm 160  183.00  & \rm 159  182.50  & \rm 158  182.30  \\
                             & \rm 1.78   0.0\% & \rm 1.36   0.2\% & \rm 1.14  10.3\% & \rm 1.15   9.8\% & \rm 1.15   9.0\% \\ \hline
(0.75,0.60,0.50)             & \rm 150  222.40  & \rm 149  172.20  & \rm 148  129.90  & \rm 147  129.80  & \rm 146  129.80  \\
                             & \rm 1.48   0.0\% & \rm 1.16   9.4\% & \rm 0.88  85.6\% & \rm 0.88  84.3\% & \rm 0.89  82.9\% \\ \hline
(0.80,0.70,0.60)             & \rm 129  173.90  & \rm 128  135.00  & \sf 127   96.30  & \sf 126   96.28  & \sf 125   94.31  \\
                             & \rm 1.35   0.5\% & \rm 1.05  32.0\% & \sf 0.76  98.1\% & \sf 0.76  97.7\% & \sf 0.75  98.1\% \\ \hline
(0.95,0.85,0.60)             & \rm 111  150.30  & \rm 110  107.20  & \sf 109   77.62  & \sf 108   77.62  & \sf 107   75.67  \\
                             & \rm 1.35   0.8\% & \rm 0.97  55.8\% & \sf 0.71  99.0\% & \sf 0.72  98.8\% & \sf 0.71  99.1\% \\ \hline
(1.00,0.90,0.60)             & \rm 105  139.20  & \rm 104  100.90  & \sf 103   70.74  & \sf 102   70.73  & \sf 101   67.50  \\
                             & \rm 1.33   1.4\% & \rm 0.97  56.7\% & \sf 0.69  99.4\% & \sf 0.69  99.2\% & \sf 0.67  99.6\% \\ \hline
($\infty$,0.90,0.65)         & \rm  92  130.00  & \rm  91   77.01  & \it  90   60.85  & \sf  89   58.66  & \sf  88   58.31  \\
                             & \rm 1.41   0.6\% & \rm 0.85  85.2\% & \it 0.68  99.2\% & \sf 0.66  99.5\% & \sf 0.66  99.4\% \\ \hline
($\infty$,$\infty$,0.65)     & \rm  78   96.09  & \rm  77   56.51  & \sf  76   49.55  & \sf  75   46.63  & \sf  74   45.94  \\
                             & \rm 1.23   8.1\% & \rm 0.73  96.2\% & \sf 0.65  99.2\% & \sf 0.62  99.6\% & \sf 0.62  99.6\% \\ \hline
($\infty$,$\infty$,$\infty$) & \rm  52   55.85  & \sf  51   25.23  & \sf  50   25.17  & \sf  49   24.11  & \sf  48   24.10  \\
                             & \rm 1.07  33.2\% & \sf 0.49  99.9\% & \sf 0.50  99.9\% & \sf 0.49  99.9\% & \sf 0.50  99.8\% \\ \hline
\end{tabular}
\vspace*{0.5in}
\doublespace
\caption{
  Degrees of freedom (DF), $\chi^2$, $\chi^2$/DF and confidence level
  for the $n^{th}$-order fit (\protect\ref{eq3}) of
  $\xi(\beta,2L)/\xi(\beta,L)$ versus $\xi(\beta,L)/L$.
  The indicated $x_{min}$ values apply to $L=8,16,32$, respectively;
  we always take $x_{min} = 0.14, 0$ for $L=64,128$.
  Our preferred fit is shown in {\em italics}\/;
   other good fits are shown in {\sf sans-serif};
   bad fits are shown in {\rm roman}.
}
\label{chisquared_xi_versus_n}
\end{center}
\end{table}

\clearpage

\singlespace
\begin{table}
\addtolength{\tabcolsep}{-1.0mm}
\footnotesize
\hspace*{-2.7cm}    
\tabcolsep 3pt      
\doublerulesep 1.5pt  
\begin{tabular}{|c||r@{\ (}r|r@{\ (}r|r@{\ (}r|r@{\ (}r|r@{\ (}r|r@{\ (}r|%
r@{\ (}r|r@{\ (}r|}
\hline
\multicolumn{1}{|c||}{$x_{min}$}
    & \multicolumn{2}{c|}{1.80} & \multicolumn{2}{c|}{2.00}
    & \multicolumn{2}{c|}{2.20} & \multicolumn{2}{c|}{2.40}
    & \multicolumn{2}{c|}{2.60} & \multicolumn{2}{c|}{2.85}
    & \multicolumn{2}{c|}{3.00} & \multicolumn{2}{c|}{3.15}  \\
\hline
 (0.70,0.55,0.45)  &  10.455  & 0.022)  &  24.903  & 0.066)  &  57.13  & 0.17)  &  129.68  & 0.41)  &  
                     290.5  & 1.0)  &  794.9  & 3.2)  &  1460  & 6)  &  2687  & 11) \\ \hline
 (0.75,0.60,0.50)  &  10.454  & 0.022)  &  24.886  & 0.071)  &  57.50  & 0.18)  &  130.83  & 0.43)  &  
                     293.0  & 1.1)  &  801.7  & 3.4)  &  1473  & 6)  &  2709  & 12) \\ \hline
\sf(0.80,0.70,0.60)  & \sf 10.450  & \sf0.021)  & \sf 24.875  & \sf0.073)  & \sf 57.41  & \sf0.22)  & \sf 130.93  & \sf0.64)  & \sf 
                     293.6  & \sf1.6)  & \sf 805.9  & \sf5.0)  & \sf 1482  & \sf9)  & \sf 2727  & \sf17) \\ \hline
\sf(0.95,0.85,0.60)  & \sf 10.451  & \sf0.021)  & \sf 24.870  & \sf0.071)  & \sf 57.40  & \sf0.21)  & \sf 130.93  & \sf0.63)  & \sf 
                     293.7  & \sf1.6)  & \sf 806.6  & \sf6.1)  & \sf 1483  & \sf12)  & \sf 2749  & \sf25) \\ \hline
\sf(1.00,0.90,0.60)  & \sf 10.450  & \sf0.022)  & \sf 24.872  & \sf0.069)  & \sf 57.40  & \sf0.21)  & \sf 130.94  & \sf0.63)  & \sf 
                     293.6  & \sf1.6)  & \sf 806.8  & \sf5.9)  & \sf 1484  & \sf12)  & \sf 2749  & \sf25) \\ \hline
\it($\infty$,0.90,0.65)  &\it10.446  &\it0.022)  &\it24.859  &\it0.072)  &\it57.40  &\it0.21)  &\it131.00  &\it0.66)  &\it
                     295.2  &\it2.1)  &\it809.6  &\it6.7)  &\it1489  &\it13)  &\it2761  &\it27) \\ \hline
($\infty$,$\infty$,0.65)  & \sf  10.447  & \sf 0.022)  & \sf  24.863  & \sf 0.074)  & \sf  57.40  & \sf 0.22)  & \sf  131.01  & \sf 0.66)  & \sf  
                    \sf295.0  & \sf 2.1)  & \sf  809.7  & \sf 6.9)  & \sf  1487  & \sf 14)  & \sf  2759  & \sf 28) \\ \hline
\sf($\infty$,$\infty$,$\infty$)  & \sf  10.454  & \sf 0.022)  & \sf  24.881  & \sf 0.074)  & \sf  57.39  & \sf 0.22)  & \sf  130.78  & \sf 0.66)  & \sf  
                     \sf295.6  & \sf 2.3)  & \sf  812.7  & \sf 9.8)  & \sf  1482  & \sf 22)  & \sf  2777  & \sf 49) \\ \hline
\end{tabular}
\par\vspace*{5mm}\par
\hspace*{-2.7cm}    
\begin{tabular}{|c||r@{\ (}r|r@{\ (}r|r@{\ (}r|r@{\ (}r|r@{\ (}r|r@{\ (}r|%
r@{\ (}r|r@{\ (}r|}
\hline
\multicolumn{1}{|c||}{$x_{min}$}
    & \multicolumn{2}{c|}{3.30} & \multicolumn{2}{c|}{3.45}
    & \multicolumn{2}{c|}{3.60} & \multicolumn{2}{c|}{3.75}
    & \multicolumn{2}{c|}{3.90} & \multicolumn{2}{c|}{4.05}
    & \multicolumn{2}{c|}{4.20} & \multicolumn{2}{c|}{4.35}  \\
\hline
 (0.70,0.55,0.45)  &   4957  &   23)  &   9117  &    46)  &   16780  &    92)  &    30959  &   182)  &    56766  &   362)  &   105205  &   707)  &   196197  &   1396)   &   360864  &   2792)  \\ \hline
 (0.75,0.60,0.50)  &   4995  &   24)  &   9199  &    47)  &   16938  &    93)  &    31258  &   185)  &    57265  &   366)  &   106093  &   736)  &   197949  &   1419)   &   363905  &   2880)  \\ \hline
\sf(0.80,0.70,0.60)  & \sf  5032  & \sf  32)  & \sf  9268  & \sf   62)  & \sf  17066  & \sf  118)  & \sf   31492  & \sf  239)  & \sf   57687  & \sf  456)  & \sf  106807  & \sf  878)  & \sf  199117  & \sf  1690)   & \sf  366159  & \sf  3309)  \\ \hline
\sf(0.95,0.85,0.60)  & \sf  5109  & \sf  49)  & \sf  9411  & \sf   92)  & \sf  17359  & \sf  178)  & \sf   32059  & \sf  346)  & \sf   58748  & \sf  650)  & \sf  108781  & \sf  1237)  & \sf  202868  & \sf  2360)   & \sf  372553  & \sf  4392)  \\ \hline
\sf(1.0,0.90,0.60)  & \sf  5110  & \sf  51)  & \sf  9365  & \sf   99)  & \sf  17299  & \sf  196)  & \sf   31816  & \sf  372)  & \sf   58308  & \sf  702)  & \sf  107789  & \sf  1312)  & \sf  200994  & \sf  2493)   & \sf  369579  & \sf  4697)  \\ \hline
\it ($\infty$,0.90,0.65)  & \it  5132  & \it  55)  & \it  9407  & \it  105)  & \it  17377  & \it  208)  & \it   31908  & \it  398)  & \it   58594  & \it  766)  & \it  108952  & \it  1452)  & \it  201796  & \it  2817)   & \it  371706  & \it  5457)  \\ \hline
\sf($\infty$,$\infty$,0.65)  & \sf 5125  & \sf 55)  & \sf 9391  & \sf 110)  & \sf 17389  & \sf 229)  & \sf  32008  & \sf 463)  & \sf  58804  & \sf 941)  & \sf 109440  & \sf 1886)  & \sf 204587  & \sf 3779)   & \sf 376704  & \sf 7722)  \\ \hline
\sf($\infty$,$\infty$,$\infty$)  & \sf 5063  & \sf 102)  & \sf 9295  & \sf 217)  & \sf 16991  & \sf 447)  & \sf  30912  & \sf 903)  & \sf  55976  & \sf 1828)  & \sf 104740  & \sf 3678)  & \sf 192664  & \sf 7358)   & \sf 359299  & \sf 14787)  \\ \hline
\end{tabular}
\vspace*{0.5in}
\doublespace
\caption{
  Estimated correlation lengths $\xi_\infty^{(2)}$ as a function of $\beta$,
  from various extrapolations.
  Error bar is one standard deviation (statistical errors only).
  All extrapolations use $s=2$ and $n=13$.
  The indicated $x_{min}$ values apply to $L=8,16,32$, respectively;
  we always take $x_{min} = 0.14, 0$ for $L=64,128$.
  Our preferred fit is shown in {\em italic}\/;
   other good fits are shown in {\sf sans-serif};
   bad fits are shown in {\rm roman}.
}
\label{table_su3_xi}
\end{table}

\clearpage


\begin{thebibliography}{99}
\bibitem{correlation_lengths}   Here $\xi^{(exp)}$ is the exponential
   correlation length (= inverse mass gap), and $\xi^{(2)}$ is the
   second-moment correlation length defined by (4.11)--(4.13) of
   Ref.~\cite{CEPS_swwo4c2}.
   Note that $\xi^{(2)}$ is well-defined in finite volume as well as in
   infinite volume;  where necessary we write $\xi^{(2)}(L)$ and
   $\xi^{(2)}_\infty$, respectively.
   In this paper, $\xi$ without a superscript denotes $\xi^{(2)}$.
\bibitem{constant_SUN} The $SU(2)$ principal chiral model is equivalent
   to the $4$-vector model; and the $1/N$ expansion of the latter model,
   evaluated at $N=4$, indicates that $C_{\xi^{(2)}}/C_{\xi^{(exp)}}
   \approx 0.9992$ \cite{oneoverN}.
   Monte Carlo data on
   the $SU(N)$ chiral model with $N=3,6,9,15,21,30$ can be found in
   \cite{Rossi_94}.
\bibitem{CPU_time} This is an ``equivalent'' CPU time; some runs on the 
   smaller lattices were performed on an IBM SP2.
\end{thebibliography}
\end{document}